\RequirePackage{lineno} 

\documentclass[aps,prl,twocolumn,superscriptaddress]{revtex4}

\usepackage{amsmath}
\usepackage{amssymb}
\usepackage{graphicx}
\usepackage{color}

\setcitestyle{super}

\pdfoutput=1

\newcommand{\eq}{\begin{equation}}
\newcommand{\eeq}{\end{equation}}
\newcommand{\ket}[1]{\left|#1\right\rangle}
\newcommand{\bra}[1]{\left\langle #1\right|}

\begin{document}

%\linenumbers

\title{Observation of a Discrete Time Crystal}

\author{J. Zhang, P. W. Hess, A. Kyprianidis, P. Becker, A. Lee, J. Smith, G. Pagano}
\affiliation{Joint Quantum Institute, University of Maryland Department of Physics and National Institute of Standards and Technology, College Park, MD  20742}

\author{I.-D. Potirniche}
\affiliation{Department of Physics, University of California Berkeley, Berkeley, CA 94720, USA}

\author{A. C. Potter}
\affiliation{Department of Physics, University of California Berkeley, Berkeley, CA 94720, USA}
\affiliation{Department of Physics, University of Texas at Austin, Austin, TX 78712, USA}

\author{A. Vishwanath}
\affiliation{Department of Physics, University of California Berkeley, Berkeley, CA 94720, USA}
\affiliation{Department of Physics, Harvard University, Cambridge, MA 02138, USA}

\author{N. Y. Yao}
\affiliation{Department of Physics, University of California Berkeley, Berkeley, CA 94720, USA}

\author{C. Monroe}
\affiliation{Joint Quantum Institute, University of Maryland Department of Physics and National Institute of Standards and Technology, College Park, MD  20742}

%J. Zhang, P. W. Hess, A. Kyprianidis, P. Becker, A. Lee, J. Smith, G. Pagano, I.-D. Potirniche, A. C. Potter, A. Vishwanath, N. Y. Yao, C. Monroe

\date{\today}

\maketitle

\textbf{Spontaneous symmetry breaking is a fundamental concept in many areas of physics, ranging from  cosmology and particle physics to condensed matter~\cite{ChaikinLubensky}. A prime example is the breaking of spatial translation symmetry, which underlies the formation of crystals and the phase transition from liquid to solid.  Analogous to crystals in space, the breaking of  translation symmetry in time and the emergence of a ``time crystal" was recently proposed~\cite{Wilczek2012,Wilczek2013a}, but later shown to be forbidden in thermal equilibrium~\cite{Bruno2013a,Bruno2013, Watanabe2015}. However, non-equilibrium Floquet systems subject to a periodic drive can exhibit persistent time-correlations at an emergent sub-harmonic frequency~\cite{Khemani2016,Else2016,Keyserlingk2016,Yao2016}. This new phase of matter has been dubbed a ``discrete time crystal" (DTC)~\cite{Yao2016,fn1}. Here, we present the first experimental observation of a discrete time crystal, in an interacting spin chain of trapped atomic ions. We apply a periodic Hamiltonian to the system under many-body localization (MBL) conditions, and observe a sub-harmonic temporal response that is robust to external perturbations. Such a time crystal opens the door for studying systems with long-range spatial-temporal correlations and novel phases of matter that emerge under intrinsically non-equilibrium conditions~\cite{Khemani2016}.} %\footnotetext{This phase is also referred to as a $\pi$-spin glass~\cite{Khemani2016} or a Floquet time crystal~\cite{Else2016}.}

%Spontaneous symmetry breaking is a fundamental concept in many areas of physics, ranging from  cosmology and particle physics to condensed matter. A prime example is the breaking of spatial translation symmetry, which underlies the formation of crystals and the phase transition from liquid to solid.  Analogous to crystals in space, the breaking of  translation symmetry in time and the emergence of a "time crystal" was recently proposed, but later shown to be forbidden in thermal equilibrium. However, non-equilibrium Floquet systems subject to a periodic drive can exhibit persistent time-correlations at an emergent sub-harmonic frequency. This new phase of matter has been dubbed a "discrete time crystal" (DTC). Here, we present the first experimental observation of a discrete time crystal, in an interacting spin chain of trapped atomic ions. We apply a periodic Hamiltonian to the system under many-body localization (MBL) conditions, and observe a sub-harmonic temporal response that is robust to external perturbations. Such a time crystal opens the door for studying systems with long-range spatial-temporal correlations and novel phases of matter that emerge under intrinsically non-equilibrium conditions.

For any symmetry in a Hamiltonian system, its spontaneous breaking in the ground state leads to a phase transition \cite{SachdevBook}.  The broken symmetry itself can assume many different forms. 
For example, the breaking of spin-rotational symmetry leads to a phase transition from paramagnetism to ferromagnetism when the temperature is brought below the Curie point.  
The breaking of spatial symmetry leads to the formation of crystals, where the continuous translation symmetry of space is replaced by a discrete one.

\begin{figure}
\includegraphics[width=\columnwidth]{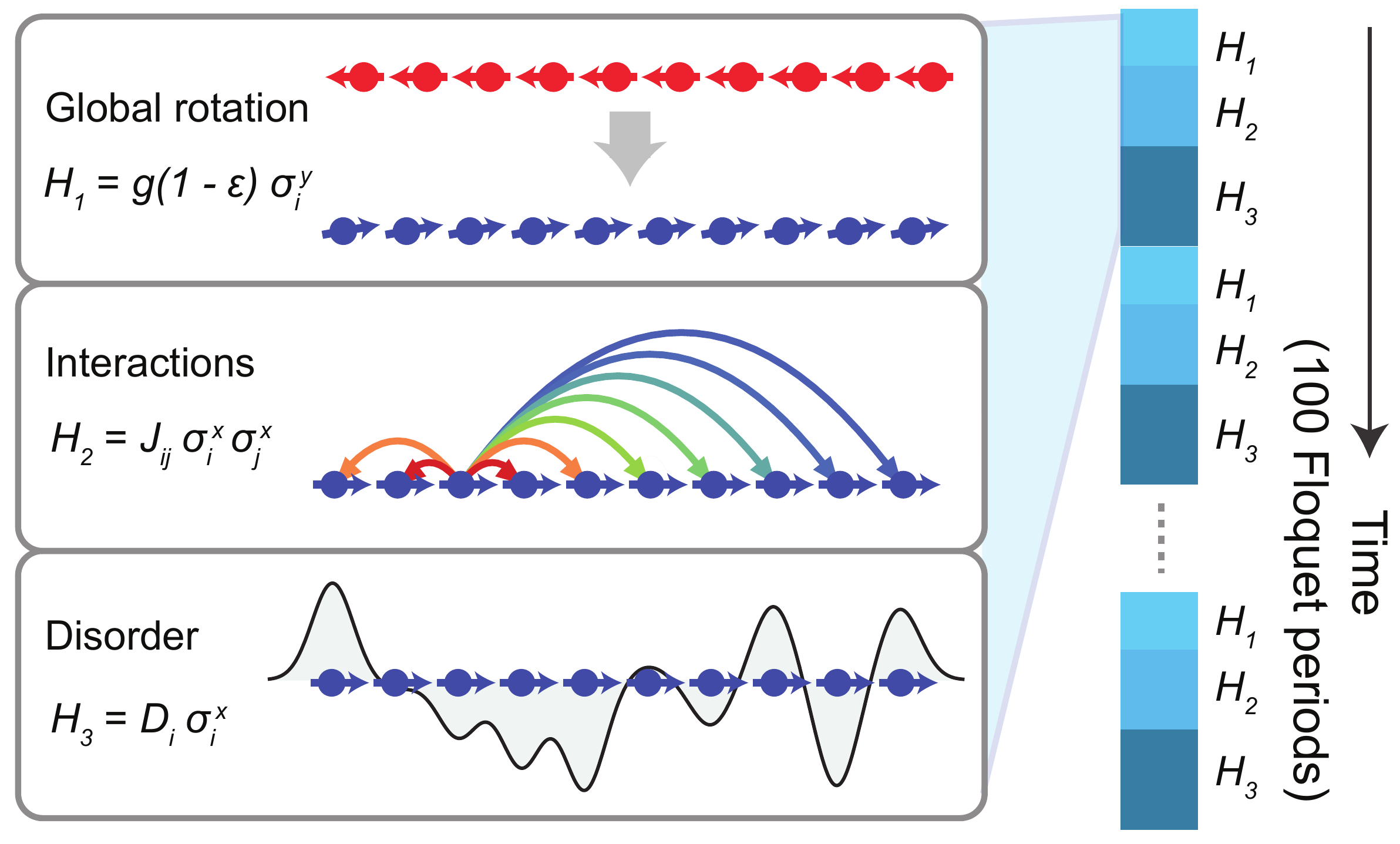}
%\linespread{1.2}
\caption{\textbf{Floquet evolution of a spin chain.} Three Hamiltonians are applied sequentially in time: a global spin flip of nearly $\pi$ $(H_1)$, long-range Ising interactions $(H_2)$, and strong disorder $(H_3)$. The system evolves for 100 Floquet periods of this sequence.}
\label{fig:Interacting}
\end{figure}

We now pose an analogous question: can the translation symmetry of time be broken?  The proposal of such a ``time crystal" \cite{Wilczek2012} for time-independent Hamiltonians has led to much discussion, with the conclusion that such structures cannot exist in the ground state or any thermal equilibrium state of a quantum mechanical system~\cite{Bruno2013, Bruno2013a, Watanabe2015}.  A simple intuitive explanation is that quantum equilibrium states have time-independent observables by construction; thus, time translation symmetry can only be spontaneously broken in non-equilibrium systems~\cite{Khemani2016,Else2016,Keyserlingk2016,Yao2016}.  In particular, the dynamics of periodically-driven Floquet systems possesses a discrete time translation symmetry governed by the drive period.  This symmetry can be further broken into ``super-lattice'' structures where  physical observables exhibit a period larger than that of the drive. Such a response is analogous to commensurate charge density waves that break the discrete translation symmetry of their underlying lattice~\cite{ChaikinLubensky}. The robust sub-harmonic synchronization of the many-body Floquet system is the essence of the discrete time crystal phase~\cite{Khemani2016,Else2016,Keyserlingk2016,Yao2016}. In a DTC, the underlying Floquet drive should generally be accompanied by strong disorder, leading to many-body localization~\cite{nandkishore2015many} and thereby preventing the quantum system from absorbing the drive energy and heating to infinite temperatures~\cite{DAlessio2014,Ponte2015,Eckardt2016,fn3}. 

\begin{figure*}
\includegraphics[width=2\columnwidth]{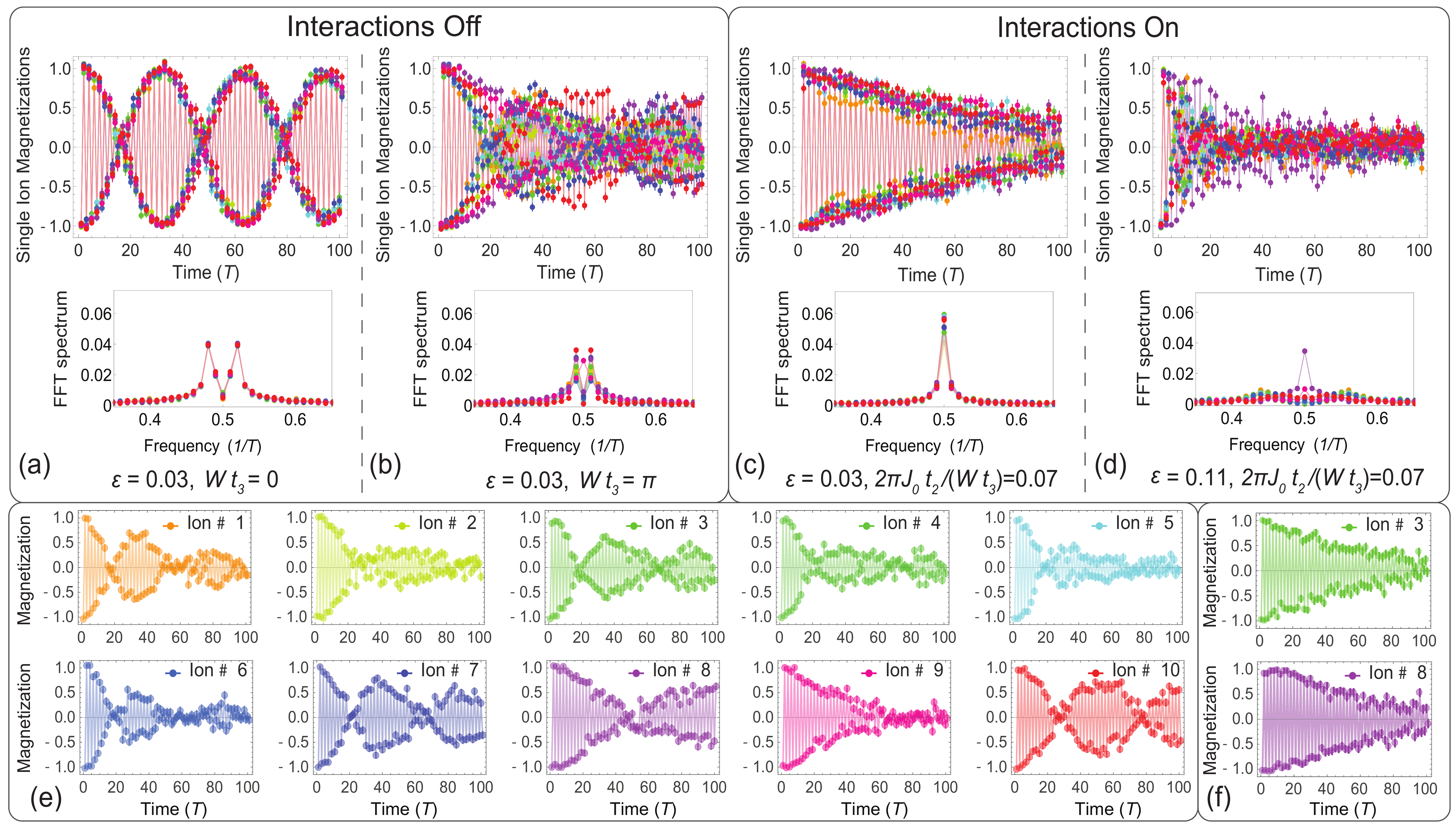}
%\linespread{1.2}
\caption{\textbf{Spontaneous breaking of discrete time translation symmetry.} Top panel: Time-evolved magnetizations of each spin $\langle \sigma_{i}^{x} (t) \rangle$ and their Fourier spectra, showing sub-harmonic response of the system to the Floquet Hamiltonian. (a) When only the $H_1$ spin flip is applied, the spins oscillate with a sub-harmonic response that beats due to the perturbation $\varepsilon = 0.03$ from perfect $\pi$-pulses, with a clear splitting in the Fourier spectrum. (b) With both the $H_1$ spin flip and the disorder $H_3$, the spins precess with various Larmor rates in the presence of different individual fields. (c) Finally, adding the spin-spin interaction term $H_2$ (shown with the largest interaction phase $J_\textrm{0}t_2 = 0.036$ rad), the spins lock to the sub-harmonic frequency of the drive period.  Here the Fourier spectrum merges into a single peak even in the face of perturbation $\varepsilon$ on the spin flip $H_1$.  (d) When the perturbation is too strong ($\varepsilon = 0.11$), we cross the boundary from the discrete time crystal into a symmetry unbroken phase~\cite{Yao2016}. Bottom panel: Individually resolved time traces. (e) Spin magnetization for all 10 spins corresponding to the case of (b). (f) Spin 3 and 8 corresponding to the case of (c). Each point is the average of 150 experimental repetitions. Error bars are computed from quantum projection noise and detection infidelities.
}
\label{fig:SpinOscillations}
\end{figure*}

Here, we report the direct observation of discrete time translation symmetry breaking and DTC formation in a spin chain of trapped atomic ions, under the influence of a periodic Floquet-MBL Hamiltonian.  We experimentally implement a quantum many-body Hamiltonian with long-range Ising interactions and disordered local effective fields, using optical control techniques~\cite{Smith2016,Lee2016}. Following the evolution through many Floquet periods, we measure the temporal correlations of the spin magnetization dynamics.

A DTC requires the ability to control the interplay between three key ingredients: strong drive, interactions, and disorder. These are reflected in the applied Floquet Hamiltonian $H$, consisting of the following three successive pieces with overall period $T=t_1+t_2+t_3 \nonumber$ (see Fig. \ref{fig:Interacting}) $(\hbar=1)$:

\begin{equation}
H = \begin{cases} 
H_1 = g(1-\varepsilon) \sum_i \sigma^y_i, & \mbox{time } t_1  \\[4pt]  
H_2 = \sum_i J_{ij} \sigma^x_i \sigma^x_{j}, \ &\mbox{time } t_2  \\[4pt]  
H_3 = \sum_{i}{D_i}\sigma_i^x \ & \mbox{time } t_3.
\end{cases}
\label{eq:model}
\end{equation}
Here, $\sigma_i^\gamma$ ($\gamma=x,y,z$) is the Pauli matrix acting on the $i^\text{th}$ spin, $g$ is the Rabi frequency with small perturbation $\varepsilon$, $J_{ij}$ is the coupling strength between spins $i$ and $j$, and $D_i$ is a site-dependent disordered potential sampled from a uniform random distribution with $D_i\in[0,W]$.

To implement the Floquet Hamiltonian, each of the effective spin-1/2 particles in the chain is encoded in the $^2$S$_{1/2}\ket{F=0,m_F=0}$ and $\ket{F=1,m_F=0}$ hyperfine `clock' states of a $^{171}$Yb$^+$ ion, denoted $\ket{\downarrow}_z$ and $\ket{\uparrow}_z$ and separated by 12.642831 GHz~\cite{Olmschenk2007}.  We store a chain of $10$ ions in a linear rf Paul trap, and apply single spin rotations using optically-driven Raman transitions between the two spin states~\cite{Hayes2010}.  Spin-spin interactions are generated by spin-dependent optical dipole forces, which give rise to a tunable long-range Ising coupling~\cite{Porras2004, QSIMPRL2009} that falls off approximately algebraically as $J_{ij} \propto J_\textrm{0}/|i-j|^\alpha$. Programmable disorder among the spins is generated by the ac Stark shift from a tightly focused laser beam that addresses each spin individually~\cite{Lee2016}. The Stark shift is an effective site-dependent $\sigma_{i}^z$ field, so we surround this operation with $\frac{\pi}{2}$-pulses to transform the field into the $x$ direction of the Bloch sphere (see Methods).  Finally, we measure the magnetization of each spin by collecting the spin-dependent fluorescence on a camera for site-resolved imaging. This allows access to the single-site magnetization, $\sigma_{i}^{\gamma}$, along any direction with a detection fidelity $>98\%$ per spin.

The unitary time evolution under a single Floquet period is
\begin{equation}
U(T) = e^{-i H_3 t_3}e^{-i H_2 t_2}e^{-i H_1 t_1}.
\label{eq2}
\end{equation}
The first evolution operator $e^{-i H_1 t_1}$ nominally flips all the spins around the $y$-axis of the Bloch sphere by an angle $2g t_1 = \pi$, but also includes a controlled perturbation in the angle, $\varepsilon\pi$, where $\varepsilon < 0.15$. This critical rotation step is susceptible to noise in the Rabi frequency (1$\%$ rms) from laser intensity instability, and also optical inhomogeneities ($<5$$\%$) across the chain due to the shape of the Raman laser beams. In order to accurately control $H_1$, we use the BB1 dynamical decoupling echo sequence~\cite{Brown2004} (see Methods) to suppress these effects, resulting in control of the rotation angle to a precision  $<0.5\%$. The second evolution operator $e^{-i H_2 t_2}$ applies the spin-spin Ising interaction, where the maximum nearest-neighbor coupling $J_\textrm{0}$ ranges from $2\pi (0.04$ kHz$)$ to $2\pi (0.25$ kHz$)$ and decays with distance at a power law exponent $\alpha=1.51$. The duration of the interaction term is set so that $J_\textrm{0}t_2<0.04$ rad of phase accumulation. The third evolution operator $e^{-i H_3 t_3}$ provides disorder to localize the system, and is programmed so that the variance of the disorder is set by $Wt_3 = \pi$. In this regime, MBL in the thermodynamic limit is expected to persist even in the presence of long-range interactions~\cite{Burin2015,Yao2014,fn2}.

To observe the DTC, we initialize the spins to the state $ \ket{\psi_0} = \left | \downarrow \right \rangle_x = \frac{1}{\sqrt{2}}(\ket{\downarrow}_z+\ket{\uparrow}_z)$ through optical pumping followed by a global $\frac{\pi}{2}$ rotation. After many periods of the above Floquet unitary Eq.(\ref{eq2}), we measure the magnetization of each spin along $x$, which gives the time-correlation function
\begin{equation}
\langle  \sigma_{i}^x(t) \rangle = \bra{\psi_0}\sigma_{i}^x(t)\sigma_{i}^x(0)\ket{\psi_0}.
\end{equation}
Figure~\ref{fig:SpinOscillations} depicts the measured spin magnetization dynamics, both in the time and the frequency domain, up to $N=100$ Floquet periods. A single Floquet period $T$ is set to a value between $74$-$75~\mu$s, depending on the parameters in the Hamiltonian. 

The global $\pi$-pulse $e^{-iH_1 t_1}$ rotates the spins roughly half way around the Bloch sphere, so that we expect a response of the system at twice the drive period $2T$, or half of the Floquet frequency.  The frequency of this sub-harmonic response in the magnetization is sensitive to the precise value of the global rotation in $H_1$ and is therefore expected to track the perturbation $\varepsilon$.  This results in coherent beats and a splitting in the Fourier spectrum by $2\varepsilon$ (Fig.~\ref{fig:SpinOscillations}(a)). When we add disorder $e^{-i H_3 t_3}$ to the Floquet period, the single spins precess at different Larmor rates (Fig.~\ref{fig:SpinOscillations}(e)) and dephase with respect to each other (Fig.~\ref{fig:SpinOscillations}(b)). Only upon adding Ising interactions $e^{-i H_2 t_2}$, and hence many-body correlations, the spin synchronization is restored (Fig.~\ref{fig:SpinOscillations}(c,f)).  

\begin{figure}[!h]
\includegraphics[width=\columnwidth]{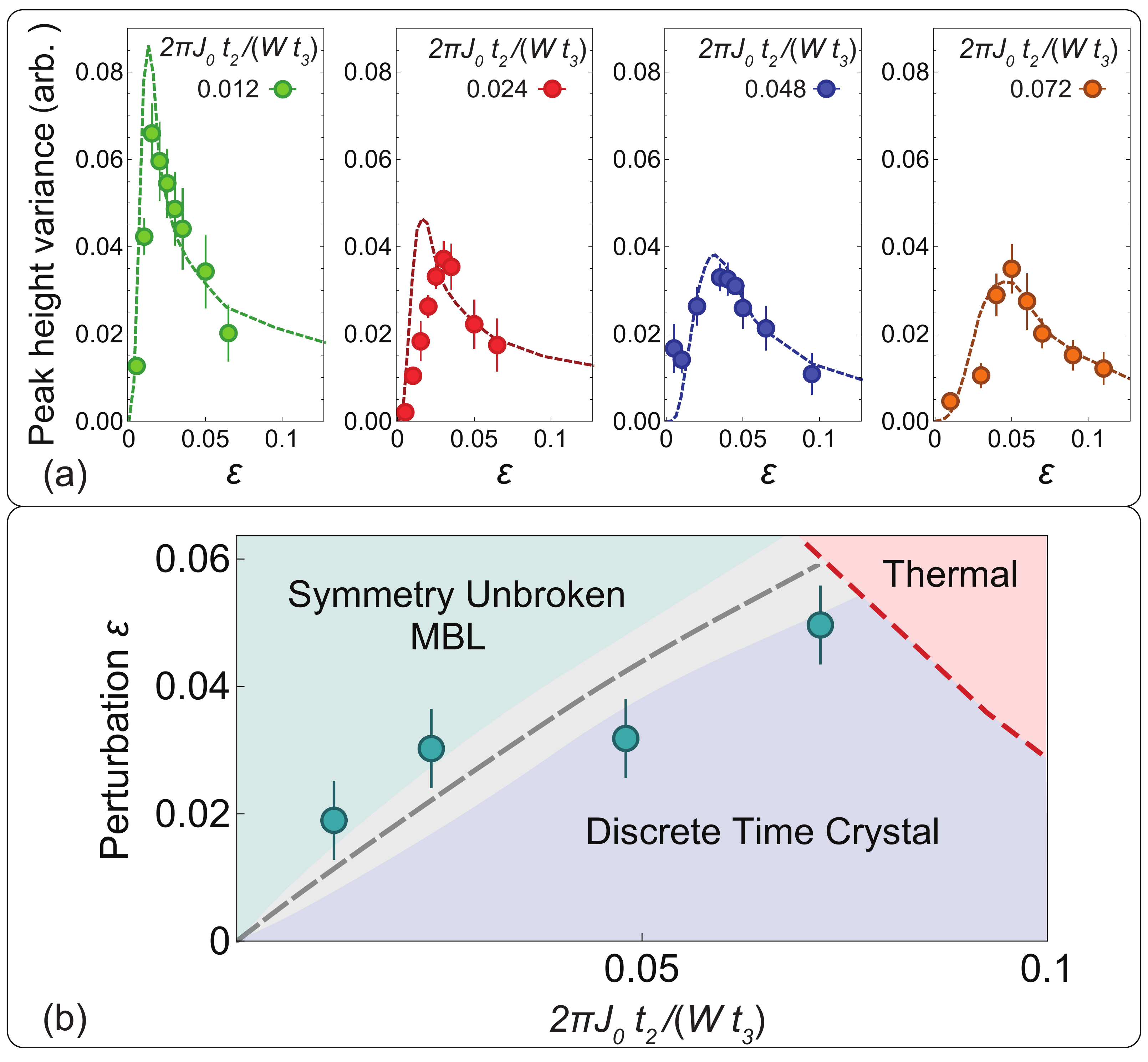}
%\linespread{1.2}
\caption{\textbf{Variance of the subharmonic peak amplitude as a signature of the DTC transition.} (a) Variances of the central peak amplitude, computed over the 10 sites and averaged over 10 instances of disorder, for four different strengths of the long-range interation term $J_\textrm{0}$.  The cross-over from symmetry unbroken state to a DTC is observed as a peak in the measured variance of the sub-harmonic system response. Dashed lines: numerical results, scaled vertically to fit the experimental data (see supplementary information). Experimental error bars are standard error of the mean. (b) Cross-over determined by a fit to the variance peak location (dots).  Dashed lines: numerically determined phase boundaries with experimental long-range coupling parameters~\cite{Yao2016}. Grey shaded region indicates 90$\%$ confidence level of the DTC to symmetry unbroken phase boundary.}
\label{fig:PeakHeightVariances}
\end{figure}

The key result is that with all of these elements, the temporal response is locked to twice the Floquet period, even in the face of perturbations to the drive in $H_1$.  This can be seen clearly as the split Fourier peaks from Fig.~\ref{fig:SpinOscillations}(b) merge into a single peak in Fig.~\ref{fig:SpinOscillations}(c).  This represents the ``rigidity" of the discrete time crystal \cite{Yao2016}, which persists under moderate perturbation strengths. However, for large $\varepsilon$, the DTC phase disappears as evinced by the decay of the sub-harmonic temporal correlations and the suppression of the central peak heights, as shown in Fig.~\ref{fig:SpinOscillations}(d). In the thermodynamic limit, these perturbations induce a phase transition from a DTC to a symmetry unbroken MBL phase~\cite{Khemani2016,Else2016,Keyserlingk2016,Yao2016}, which is rounded into a cross-over in finite size systems. 

The phase boundary is defined by the competition between the drive perturbation $\varepsilon$ and strength of the interactions $J_\textrm{0}$. We probe this boundary by measuring the variance of the sub-harmonic spectral peak height, computed over the 10 sites and averaged over 10 instances of disorder. Figure~\ref{fig:PeakHeightVariances}(a) shows the variances as a function of the perturbation $\varepsilon$, for four different interaction strengths. As we increase $\varepsilon$, the variance growth distinctively captures the onset of the transition, with increased fluctuations signaling the crossing of the phase boundary. When the perturbations are too large, the crystal ``melts". Figure~\ref{fig:PeakHeightVariances}(b) shows the fitted centers of the variance curve, on top of numerically computed phase boundaries with experimental parameters. The measurements are in agreement with the expected DTC to time crystal ``melting'' boundary, which displays approximately linear dependence on the perturbation strength in the limit of small interactions~\cite{Yao2016}.

\begin{figure}[!h]
\includegraphics[width=\columnwidth]{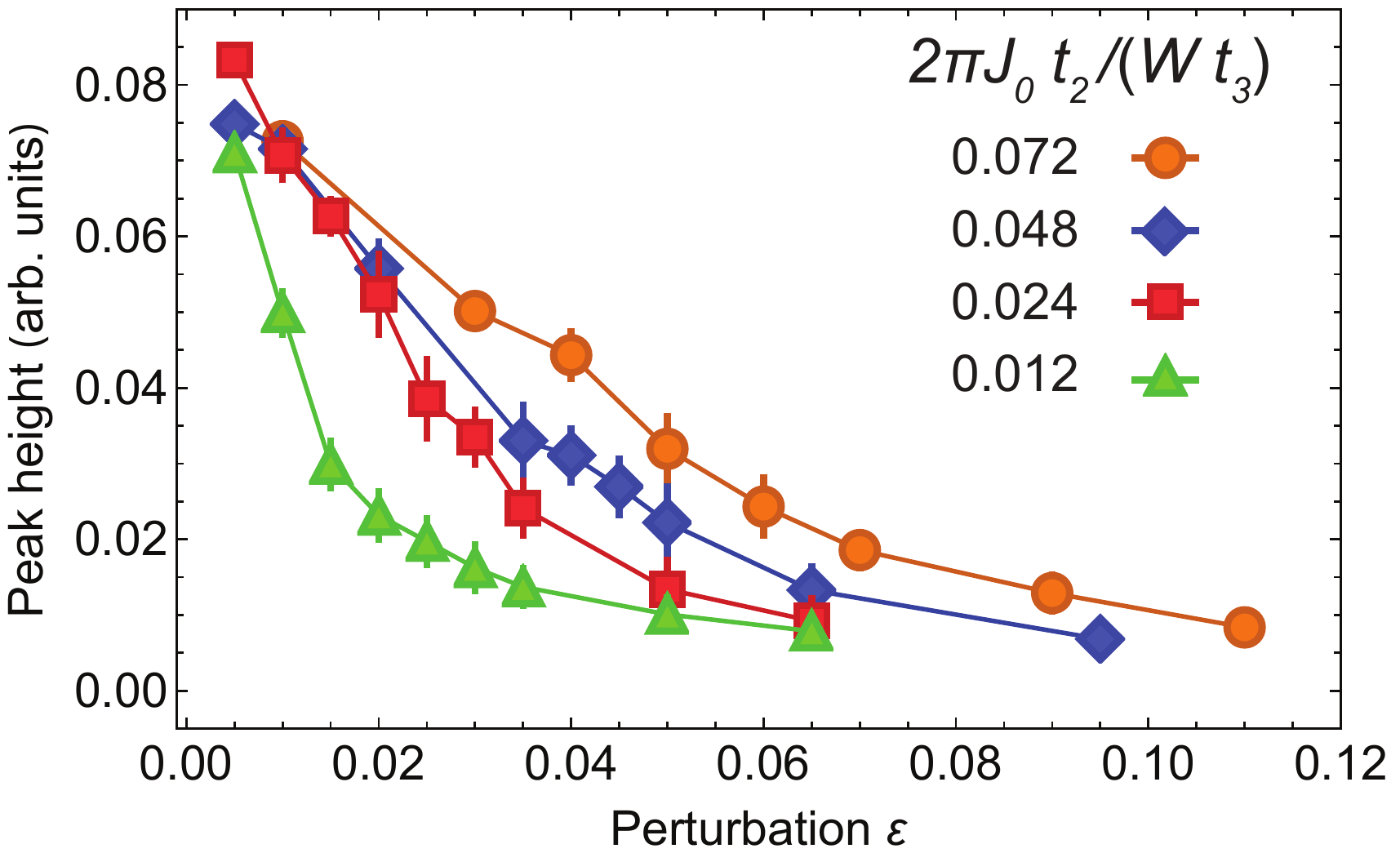}
%\linespread{1.2}
\caption{
\textbf{Subharmonic peak amplitude as a function of the drive perturbation.} Central subharmonic peak amplitude in the Fourier spectrum as a function of the perturbation $\varepsilon$, averaged over the 10 sites and 10 disorder instances, for four different interaction strengths. Solid lines are guides to the eye. The height decreases across the phase boundary and eventually diminishes as the single peak is split into two. Error bars: 1 s.d.}
\label{fig:PeakHeights}
\end{figure}

Figure~\ref{fig:PeakHeights} illustrates the amplitude of the subharmonic peak as a function of $\varepsilon$, for the four different applied interaction strengths. 
The amplitude drops off with larger perturbations, and this slope is steeper as we decrease the interaction strength.
The sub-harmonic peak amplitude observable in Fig.~\ref{fig:PeakHeights} can serve as an auxilary probe of the rigidity. It is expected to be an order parameter for the DTC phase and thus, to scale similarly to  the mutual information between spins~\cite{Else2016, Yao2016}. This connection also provides insight into the Floquet many-body quantum dynamics, and in particular to the correlations or entanglement that underly the DTC phase. Indeed, the eigenstates of the entire Floquet unitary are expected to resemble GHZ or spin-``Schr\"odinger Cat" states \cite{Else2016,Else2016a}. The initial product state in the experiment can be written as a superposition of two cat states: 
$\left | \downarrow \downarrow \cdots \downarrow \right \rangle_x = \frac{1}{\sqrt{2}}(\left | \phi_+ \right \rangle + \left | \phi_- \right \rangle)$, where $\left | \phi_\pm \right \rangle = \frac{1}{\sqrt{2}}(\left | \downarrow \downarrow \cdots \downarrow \right \rangle_x \pm \left | \uparrow \uparrow \cdots \uparrow \right \rangle_x)$. These two states evolve at different rates corresponding to their respective quasi-energies, giving rise to the sub-harmonic periodic oscillations of physical observables. Such oscillations are expected to persist at increasingly long times as the system size increases~\cite{Khemani2016,Else2016, Yao2016}. %Of course, as the system grows from finite size to the thermodyamic limit, we expect sharpening of the transition from a cross-over to a phase transition~\cite{Else2016,Yao2016}.

In summary, we present the first experimental observation of discrete time translation symmetry breaking into a DTC.  We measure persistent oscillations and synchronizations of interacting spins in a chain and show that the discrete time crystal is rigid, or robust to perturbations in the drive. Our Floquet-MBL system with long-range interactions provides an ideal testbed for out-of-equilibrium quantum dynamics and the study of novel phases of matter that  exist only in a Floquet setting~\cite{Khemani2016,Else2016,Keyserlingk2016,Yao2016,Keyserlingk2016a,Bordia2016}. Such phases can also exhibit topological order~\cite{Keyserlingk2016a,Keyserlingk2016b,Else2016b,Potter2016,Roy2016} and can be used for various quantum information tasks, such as implementing a robust quantum memory~\cite{Bahri2016}.

\section{Acknowledgements}
We acknowledge useful discussions with  Mike Zaletel and Dan Stamper-Kurn.  This work is supported by the ARO Atomic and Molecular Physics Program, the AFOSR MURI on Quantum Measurement and Verification, the IARPA LogiQ program, the IC Postdoctoral Research Fellowship Program, the NSF Physics Frontier Center at JQI, and the Miller Institute for Basic Research in Science. A. V. was supported by the AFOSR MURI grant FA9550- 14-1-0035 and Simons Investigator Program.

%\bibliographystyle{naturemag}

%\bibliography{timecrystal}

%\footnote{This phase is also referred to as a $\pi$-spin glass~\cite{Khemani2016} or a Floquet time crystal~\cite{Else2016}.}

%\footnote{We note that under certain conditions, time-crystal dynamics can persist for rather long times even in the absence of localization before ultimately being destroyed by thermalization \cite{Else2016a}.

%\footnote{There has been recent work~\cite{DeRoeck2016} questioning the stability of MBL with anything longer-range than exponential-in-distance interaction. However, the proposed mechanism is not relevant on  experimentally accessible length- and time- scales.}

\section{METHODS SUMMARY}

\subsection{Dynamical decoupling sequence.}

We use a pair of Raman laser beams globally illuminating the entire ion chain to drive qubit rotations~\cite{Hayes2010}. The ion chain has 25 $\mu m$ length, and we shape the beams to have 200 $\mu m$ full width half maximum along the ion chain, resulting in $\sim5\%$ intensity inhomogeneity. When a fixed duration is set for $H_1$ in Eq. 1 of the main text, the time dependent magnetization for different ions acquire different evolution frequencies, resulting in the net magnetization of the system dephasing after about 10 $\pi$-pulses.
In addition, the Raman laser has rms intensity noise of about 1$\%$, which restricts the spin-rotation coherence to only about 30 $\pi$-pulses (80$\%$ contrast). 

To mitigate these imperfections, we employ a BB1 dynamical decoupling pulse sequence for the drive unitary $U_1$ (written for each spin $i$):
\begin{equation}
U_1 (\varepsilon) = e^{-i H_1 t_1} =e^{-i\frac{\pi}{2}\sigma_i^{\theta}} e^{-i\pi\sigma_i^{3\theta}} e^{-i\frac{\pi}{2}\sigma_i^{\theta}} e^{-i\frac{\pi}{2}(1-\varepsilon)\sigma_i^{y}} , \nonumber \\
\end{equation}
where in addition to the perturbed $\pi$ rotation $e^{-i\frac{\pi}{2}(1-\varepsilon)\sigma_i^{y}}$, three additional rotations are applied: a $\pi$-pulse along an angle $\theta = \textrm{arccos}(\frac{-\pi}{4\pi})$, a $2\pi$-pulse along $3\theta$, and another $\pi$-pulse along $\theta$, where the axes of these additional rotations are in the $x$-$y$ plane of the Bloch sphere with the specified angle referenced to the $x$-axis. In this way, any deviation in the original rotation from the desired value of $\pi(1-\varepsilon)$ is reduced to second order. See Ref.~\cite{Brown2004} for detailed discussions. 

\subsection{Generating the effective Ising Hamiltonian}

We generate spin-spin interactions by applying spin-dependent optical dipole forces to ions confined in a 3-layer linear Paul trap with a 4.8 MHz transverse center-of-mass motional frequency. Two off-resonant laser beams with a wavevector difference $\Delta \vec{k}$ along a principal axis of transverse motion globally address the ions and drive stimulated Raman transitions. The two beams contain a pair of beatnote frequencies symmetrically detuned from the spin transition frequency by an amount $\mu$, comparable to the transverse motional mode frequencies. In the Lamb-Dicke regime, this results in the Ising-type Hamiltonian in Eq. (1)~\cite{Porras2004,QSIMPRL2009} with
\begin{equation}
\label{eqn:Jij}
J_{i,j}=\Omega^2\omega_R \sum_{m=1}^N \frac{b_{i,m}b_{j,m}}{\mu^2-\omega_m^2},
\end{equation}
where $\Omega$ is the global Rabi frequency, \mbox{$\omega_R=\hbar\Delta k^2/(2M)$} is the recoil frequency, $b_{i,m}$ is the normal-mode matrix, and $\omega_m$ are the transverse mode frequencies. The coupling profile may be approximated as a power-law decay $J_{i,j}\approx J_\textrm{0}/|i-j|^\alpha$, where in principle $\alpha$ can be tuned between 0 and 3 by varying the laser detuning $\mu$ or the trap frequencies $\omega_m$. In this work, $\alpha$ is fixed at 1.51 by setting the axial trapping frequency to be 0.44 MHz, and Raman beatnode detuning to be 155 kHz.

\subsection{Apply disorder in the axial direction.}

We apply the strong random disordered field with a fourth-order ac Stark shift~\cite{Lee2016}, which is naturally an effective $\sigma^z_i$ operator. To transform this into a $\sigma_i^x$ operator, we apply additional $\pi/2$ rotations. Hence the third term in the Floquet evolution $U_3$ (written for each spin $i$) is also a composite sequence:
\begin{equation}
U_3 = e^{-i H_3 t_3} = e^{i \frac{\pi}{4} \sigma_i^{y}} \ e^{-i D_i\sigma_{i}^{z} t_3} e^{-i \frac{\pi}{4} \sigma_i^{y}} \nonumber\\
= \ e^{-i D_i\sigma_{i}^{x} t_3}.
\end{equation}

\section{METHODS}

\subsection{Experimental time sequence.}

\begin{figure*}[!h]
\includegraphics*[width=2\columnwidth]{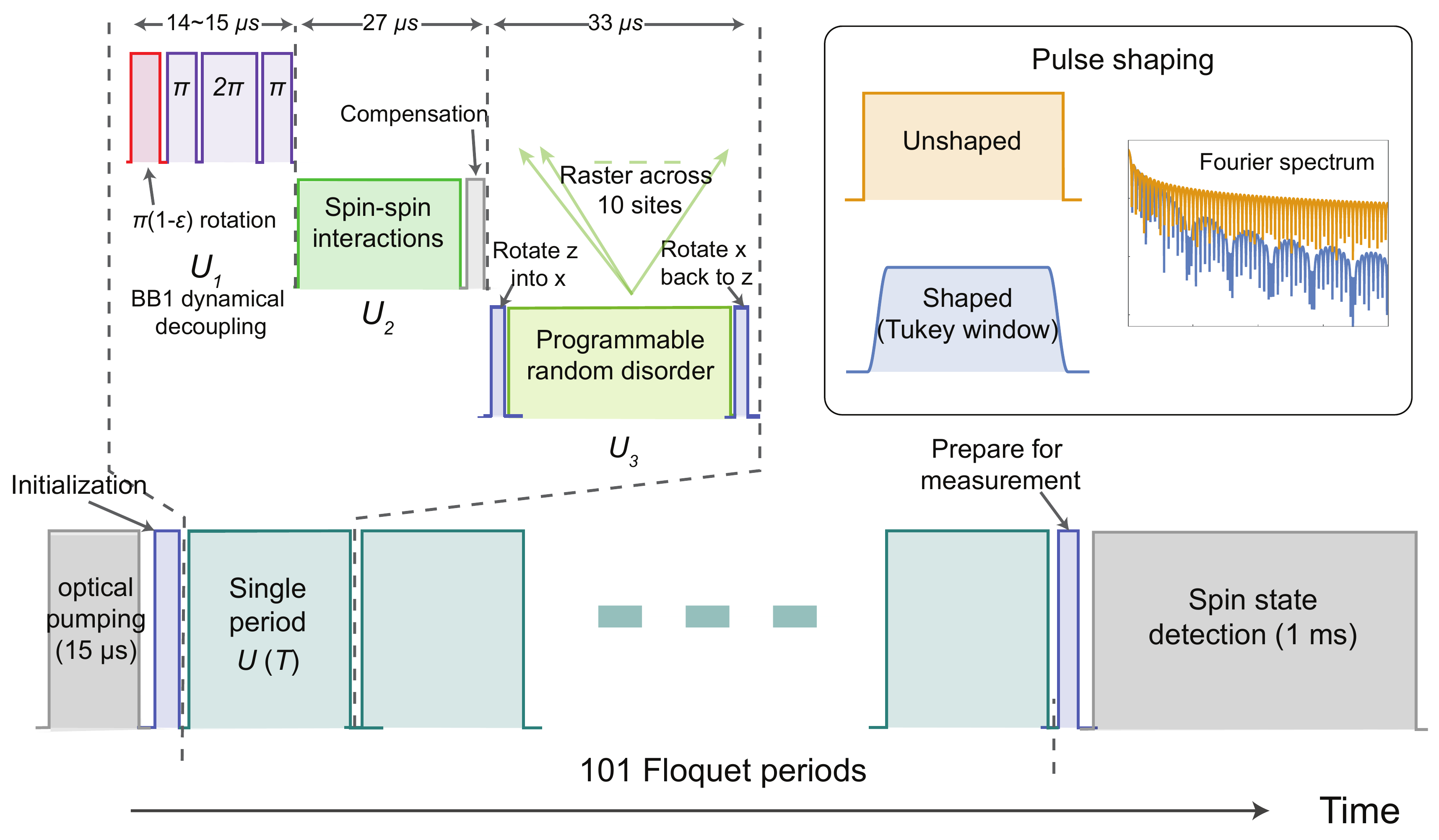}
%\linespread{1.2}
\caption{
\textbf{Experimental pulse sequence.} See text for detailed explanations.}
\label{sequence}
\end{figure*}

The Floquet time evolution is realized using the timing sequence illustrated in Fig.~\ref{sequence}.

The chain of 10 trapped ions is initialized in the ground motional state of their center of mass motion using doppler cooling and sideband cooling (not shown). Optical pumping prepares the ions in the $\ket{\downarrow}_z$ state. We then globally rotate each spin vector onto $\ket{\downarrow}_x$ by performing a $\frac{\pi}{2}$-pulse around the $y$-axis (``Initialization'' in Fig.~\ref{sequence}).

The $H_1$ Hamiltonian (perturbed $\pi$-pulses) lasts 14-15 $\mu$s depending on $\varepsilon$ and it consists of a four pulse BB1 sequence as described in the methods summary. Our carrier Rabi frequency is set such that we perform a $\pi$-pulse in less than 3 $\mu$s. Including the three compensating pulses, the BB1 sequence requires 5 $\pi$-pulse times to implement $U_1(\varepsilon)$.

 %$e^{-i\pi/2 (1-\epsilon)\sigma_i^y}$  (a perturbed $\pi-\epsilon$ pulse around the $y$ axis), $e^{-i\frac{\pi}{2} \sigma_i^\theta}$ (a $\pi$ pulse around $\theta=\arccos\frac{-\pi}{4\pi}$), $e^{-i\pi \sigma_i^{3\theta}}$ (a $2\pi$ pulse around $3\theta$) and $e^{-i\frac{\pi}{2} \sigma_i^\theta}$ (a $\pi$ pulse around $\theta$).

The $H_2$ Ising Hamiltonian is applied for 25 $\mu$s (``Spin-spin interactions'' in Fig.~\ref{sequence}). The pulse time was sufficiently long that with the pulse shaping described below the effects of the finite pulse time spectral broadening were largely reduced. To compensate for the residual off-resonant carrier drive, we apply a small amplitude transverse field (``Compensation'' in Fig.~\ref{sequence}) for 2 $\mu$s.

For the $H_3$ disorder Hamiltonian, we apply $\sum_{i=1}^N D_i\sigma_i^z$ (``Strong random disorder'' in Fig.~\ref{sequence}) generated by Stark shifts as described in the Methods summary. This is sandwiched in-between rotations around the  $y$-axis to convert this into disorder in $\sigma_i^x$. After up-to 100 applications of the Floquet evolution, we rotate the state back to the $z$-axis (``Prepare for measurement'' in Fig.~\ref{sequence}) and detect the spin state $\ket{\uparrow}_z$ or $\ket{\downarrow}_z$ using spin-dependent fluorescence.

\subsection{Pulse shaping for suppressing off-resonant excitations.}

The optical control fields for generating $H_1$, $H_2$ and $H_3$ are amplitude modulated using acousto-optic modulators (AOMs) to generate the evolution operators. If the RF drive to these modulators is applied as a square pulse will be significantly broadened in the Fourier domain due to fast rise and fall times at the edges (100 ns). As the pulse duration decreases, the width of the spectral broadening will expand (see Fig~\ref{sequence} inset). The components of the evolution operator must be sufficiently short in order to evolve 100 Floquet periods within a decoherence time of $<8$ ms. 

This spectral broadening is problematic when generating the interaction Hamiltonian $H_2$ due to off-resonant driving. The spin-spin interactions are applied using beatnote frequencies detuned 4.8~MHz from the carrier transition and 155~kHz from the sidebands. A broad pulse in the frequency domain can drive either the qubit hyperfine transition at the carrier frequency or phonons via sideband transitions.

Similar issues occur while we apply the disordered field in $H_3$. The fourth-order ac Stark shift is generated from a frequency comb that has a closest beatnote which is 23 MHz away from hyperfine and Zeeman transitions. To apply large average Stark shifts (33 kHz max.) across the 10 ions, we raster the laser beam once within a single cycle (30 $\mu$s), for a single-ion pulse duration of 3$\mu$s. The fast rastering also produces off-resonant carrier driving that resemble $\sigma_{i}^y$ fields. 

We mitigate both these effects by shaping the pulses with a 25$\%$ ``Tukey" window~, a cosine tapered function for the rise and fall. This largely removes off-resonant terms in all Hamiltonian chapters, while minimizing the reduction in pulse area (80$\%$). We carefully characterize any residual effects in $H_2$ with a single ion where no interaction dynamics are present, and apply a small compensation field in $\sigma_i^y$ to cancel residual effects. We deduce an upper limit of 0.3$\%$ (relative to $H_1$) on the residuals from the envelopes of the dynamically decoupled $\pi$ pulse sequence.

\subsection{Emergence of time crystal stabilized by the interactions and the disorder.}

\begin{figure*}[!h]
\includegraphics*[width=1.5\columnwidth]{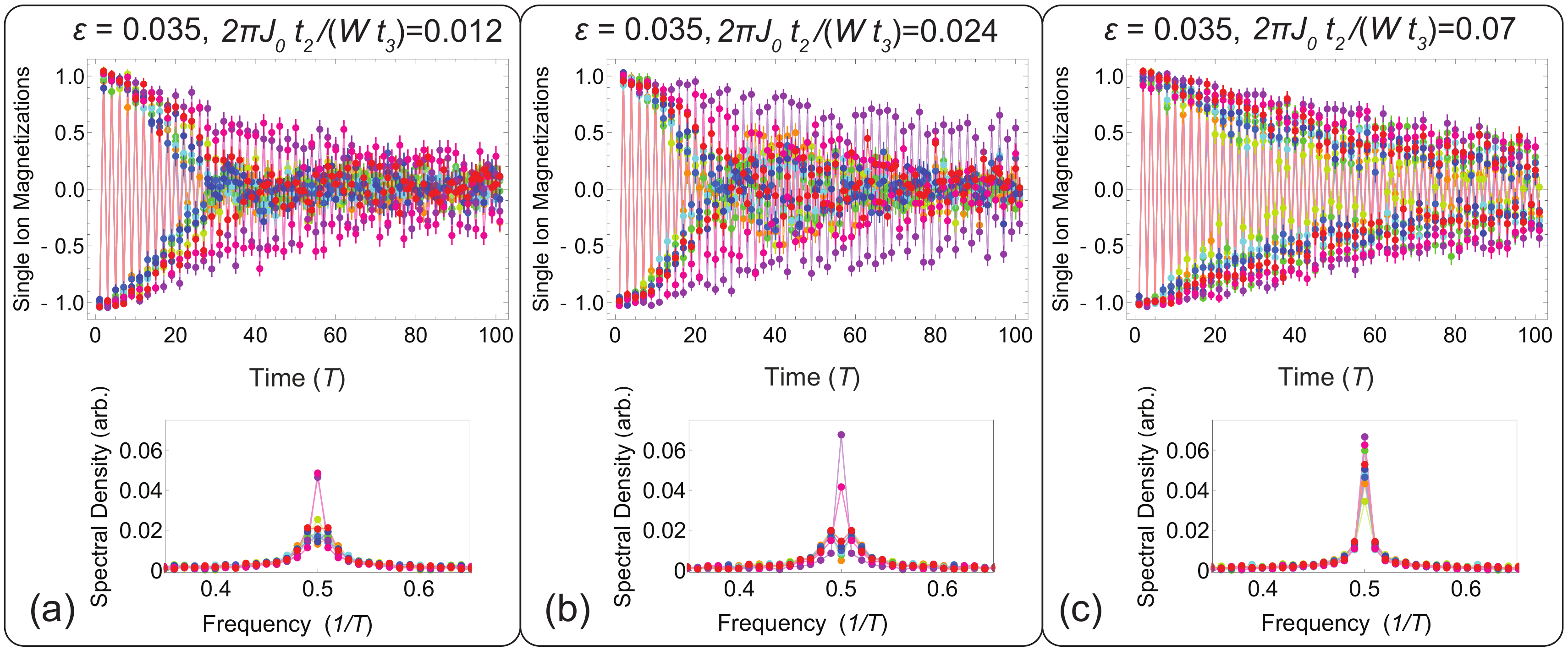}
%\linespread{1.2}
\caption{
\textbf{Build up of a discrete time-crystal.}
From (a) to (b) and (c), we fix the disorder instance and the perturbation, while gradually increase the interactions. The temporal oscillations are synchronized with increasing interactions, and the Fourier sub-harmonic peak is enhanced.}
\label{buildup}
\end{figure*}

In the main text we have shown different scenarios which occur when we modify sub-sections of the total Floquet evolution operator. Here we first expand the results shown in Fig.2 of the main text by highlighting the effect of the interactions. Figure~\ref{buildup} shows individual magnetization evolutions with the same $\varepsilon$ and instance of disorder $\{D_i\}$ but with increasing interaction $J_0$, from left to right. This shows how, all else being equal, as we turn up the interaction strength the synchronization across the ions gradually builds up signaling the formation of a discrete time crystal.

\begin{figure*}[!h]
\includegraphics*[width=2\columnwidth]{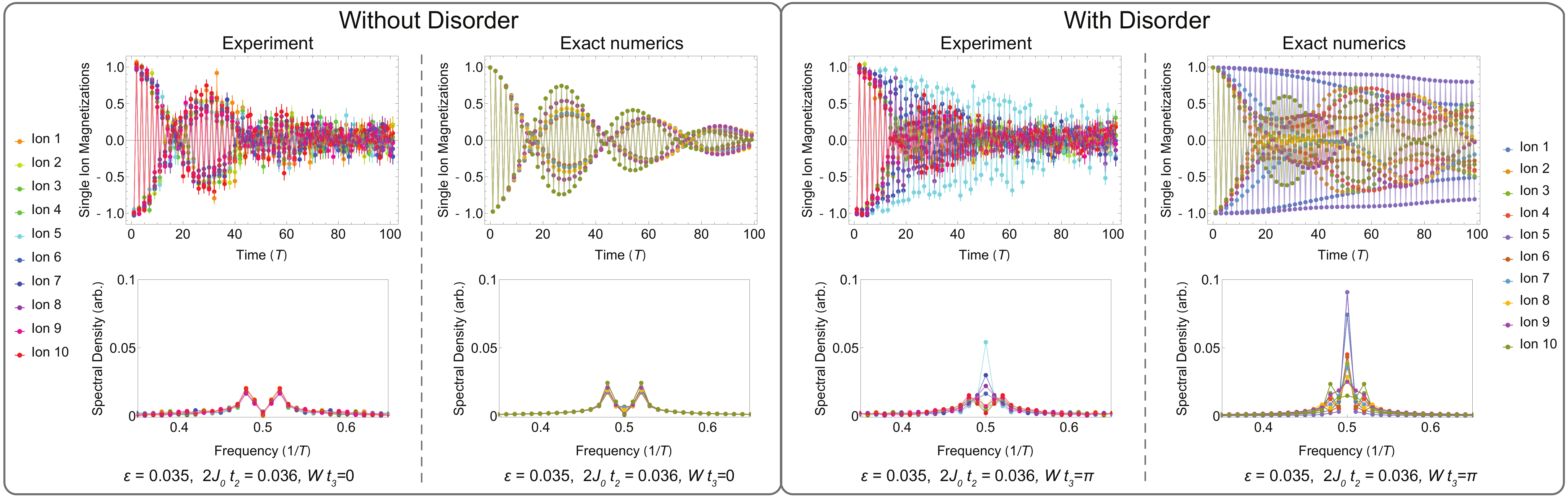}
%\linespread{1.2}
\caption{
\textbf{Comparing ions dynamics with and without disorder.} (a) Without disorder, the interaction suppresses the beating. Left: experimental data; right: exact numerics  calculated under the Floquet time-evolution. (b) With disorder, the time-crystal is more stable. Left: experimnetal result for one single instance; right: numerical simulations.}
\label{Disorder_no_yes}
\end{figure*}

In a complementary fashion, Fig.~\ref{Disorder_no_yes} addresses the role of disorder in stabilizing the time crystal formation. In particular we compare the magnetization dynamics, both in the time and frequency domain, including or removing the disorder chapter from the total Floquet evolution. We show on the side the corresponding exact numerics calculated by applying $U(T)^{N}$ on the initial state $\ket{\psi_0}$ with the measured experimental parameters. Figure~\ref{Disorder_no_yes}(a) shows that, with no disorder, the ions coherent dynamics tracks the perturbation $\varepsilon$, which results in coherent beats in agreement with our exact results. On the other hand, Figure~\ref{Disorder_no_yes}(b) shows that, all else being equal, adding the disorder chapter locks the subharmonic response of all the ions. Although the numerics is qualitatively in agreement with the experimental data, nevertheless we observe a decay in the magnetization that cannot be explained by our numerical unitary simulations.  

This damping can be due to two possible sources: one is the residual off-resonant drive of the disorder chapter which is not totally eliminated by the pulse shaping (see pulse-shaping section above). This small residual effect can behave like residual $\sigma_i^x$ and $\sigma_i^y$ terms, or coupling between the clock spin states and the Zeeman states $\ket{F=1,m_F=\pm1}$, resulting in a decay in the coherent oscillations. This effect varies across the different disorder instances and the different interaction strengths and leads to an overall decrease of the Fourier subharmonic peak height and of its variance  with respect to what is expected from the theory (see Fig.\ref{Disorder_no_yes}(b)).
To take this effect into account in the theory, we perform a least squares fit of the amplitude of the four theory curves to the experimental data. With this procedure we obtain the scaling factors (0.56, 0.53, 0.51, 0.78) for the interaction strengths $J_0t_2/(W t_3) =(0.72, 0.48, 0.24, 0.12)$ respectively, which are consistent with the decays discussed above.
%The theory curves are calculated applying the unitary evolution operator in Eq.~(2) to the initial state $\ket{\psi_0}$ for $N=100$ Floquet periods, averaging over 1000 disorder instances and extracting the peak height variance as a function of the perturbation $\varepsilon$.

\subsection{Data analysis and fitting procedure}

The peak height variance data in Fig.~3(a) are fit to a lineshape in order to extract the cross-over transition boundary $\varepsilon_p$ in Fig.~3(b). We use the following phenomenologial lineshape, a Lorentzian of the $\log_{10}(\epsilon)$
\begin{equation}
F(\varepsilon) = A \frac{1}{1+\left(\frac{\log_{10}(\varepsilon/\varepsilon_p)}{\gamma}\right)^2} + B
\label{eq:FitLineshape}
\end{equation}

A statistical error is extracted from a weighted non-linear fit to the data, which yields a fractional standard error bar of a few percent. The error in the peak height is limited by systematic error in the finite number of instances we realized in the experiment. For each value of $J_0$ and $\varepsilon$ we average over the same 10 instance of disorder. These 10 instances are averaged and fit to $F(\varepsilon)$. 

\begin{figure*}[!h]
\centering
\includegraphics*[width=2\columnwidth]{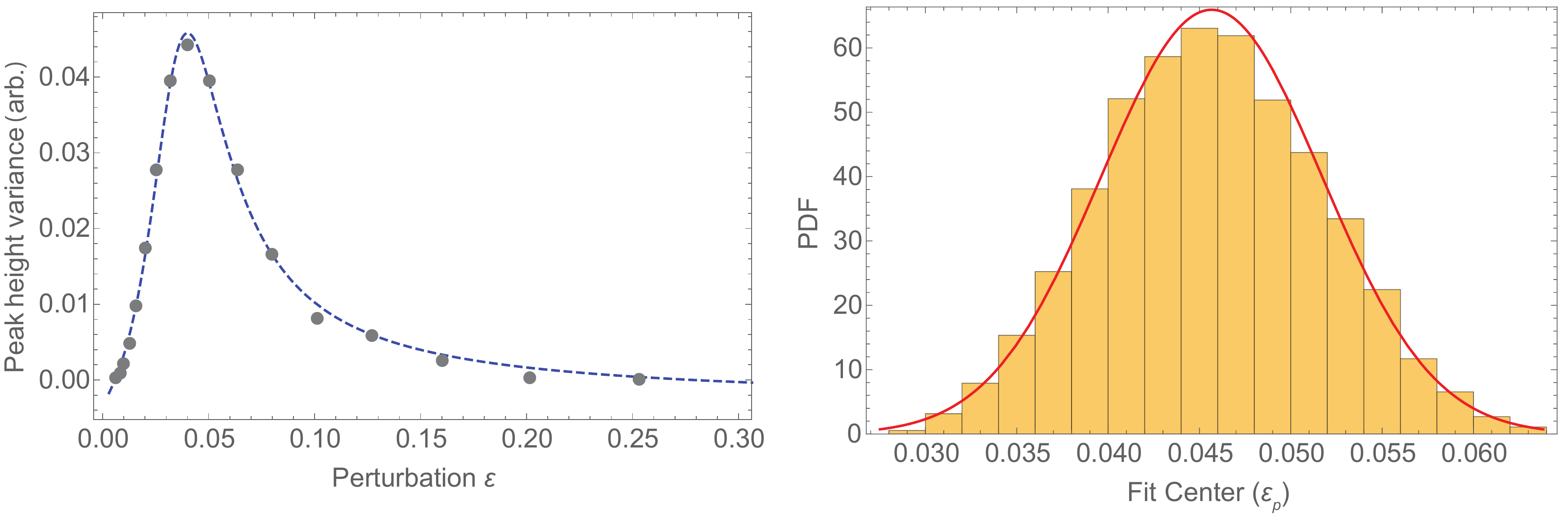}
%\includegraphics*[scale=0.5]{BootstrapDistribution.pdf}
%\linespread{1.2}
\caption{
\textbf{A random sampling from numerical evolutions.} Averages of 10 disorder instances from numerical evolution under $H$ for $2\pi J_0 t_2 / (W t_3) = 0.072$. \textbf{Left:} An example random numerical dataset (points) and fit to Methods Eq.~\ref{eq:FitLineshape} (dashed line). \textbf{Right:}  The normalized probability distribution (PDF) of peak fit centers $\varepsilon_p$ is shown in yellow, and a normal distribution is overlaid in red. The normal distribution is fit using only the mean and standard deviation of the sample, showing excellent agreement with Gaussian statistics. For this value of $J_0$ the mean $\bar{\varepsilon}_p=0.046$ and the standard deviation $\sigma_{\varepsilon_p} = 0.006$.}
\label{fig:BootDist}
\end{figure*}

In order to estimate the error due to the finite disorder instances, we perform random sampling from a numerical dataset of 100 instances of disorder. Sampling 10 of these and fitting the peak height variance data to $F(\varepsilon)$ yields a Gaussian distribution of extracted peak centers $\epsilon_p$ over 10000 repetitions (Fig.~\ref{fig:BootDist}). We take the one standard deviation (67$\%$) confidence interval in the sample as the systematic uncertainty in the fit, which yield a fractional error of $\sim 15\%$. This systematic uncertainty dominates over the statistical, so we apply error bars in Fig.~3(b) equal to this computed finite sampling error.

%\bibliographystyle{naturemag.bst}
%\bibliography{timeCrystal}

\end{document}